\documentclass[letterpaper, 10 pt, conference]{ieeeconf}%
\usepackage{bm}
\usepackage{amssymb}
\usepackage{epsfig}
\usepackage{amsmath}
\usepackage{xcolor}
\usepackage{graphicx}
\usepackage{epstopdf}
\usepackage{amsfonts}
\usepackage{indentfirst}%
\usepackage[ruled]{algorithm2e}
\usepackage{microtype}
\usepackage{booktabs} 
\usepackage{array}

\newif\ifmarginnotes
\marginnotestrue 
\ifmarginnotes
    \newcommand{\RAnote}[1]{%
    \marginpar{\tiny\color{red}#1}}
\else
    \newcommand{\RAnote}[1]{}
\fi

\providecommand{\U}[1]{\protect\rule{.1in}{.1in}}
\newtheorem{problem}{\textbf{Problem}}
\newtheorem{definition}{\textbf{Definition}}

\newtheorem{theorem}{\rm\textbf{Theorem}}
\newtheorem{lemma}{\rm\textbf{Lemma}}

\newtheorem{remark}{\rm\textbf{Remark}}
\IEEEoverridecommandlockouts
\begin{document}

\title{{\LARGE \textbf{Finite-time Convergent Control Barrier Functions with Feasibility Guarantees
}}}
\author{Anni Li, Yingqing Chen, Christos G. Cassandras and Wei Xiao\thanks{Anni Li is with the Electrical and Computer Engineering department at UNC-Charlotte. \texttt{{\small ali20@charlotte.edu}} }
\thanks{Yingqing Chen, Christos G. Cassandras are with Division of Systems Engineering at Boston University. \texttt{{\small \{yqchenn,cgc\}@bu.edu}}.}
\thanks{Wei Xiao is with the robotics engineering department at WPI and with CSAIL, MIT. \texttt{{\small weixy@mit.edu}}.}}
\maketitle

\begin{abstract}
This paper studies the problem of finite-time convergence to a prescribed safe set for nonlinear systems whose initial states violate the safety constraints. Existing Control Lyapunov-Barrier Functions (CLBFs) can enforce recovery to the safe set but may suffer from the issue of chattering and they do not explicitly consider control bounds. 
To address these limitations, we propose a new Control Barrier Function (CBF) formulation that guarantees finite-time convergence to the safe set while ensuring feasibility under control constraints. 
Specifically, we strengthen the initially violated safety constraint by introducing a parameter which enables the exploitation of the asymptotic property of a CBF to converge to the safe set in finite time. 
Furthermore, the conditions for the existence of such a CBF under control bounds to achieve finite-time convergence are derived via reachability analysis and constraint comparison, providing a systematic approach for parameter design.
A case study on 2D obstacle avoidance is presented to demonstrate the effectiveness and advantages of the proposed method.
\end{abstract}

\thispagestyle{empty} \pagestyle{empty}



\section{INTRODUCTION}

\label{sec:intro}

Constrained optimal control problems subject to safety requirements are central 
to cyber-physical systems and multi-agent systems with safety-critical autonomous agents.
Ensuring strict constraint satisfaction in such dynamic systems is therefore a key challenge,
for which Barrier Functions (BFs) have emerged as an effective tool in solving optimization problems \cite{Boyd2004}. 

In control systems, BFs are Lyapunov-like functions \cite{Tee2009},
\cite{Wieland2007}. They have been used to prove set
invariance \cite{Aubin2009}, \cite{Prajna2007}, \cite{Wisniewski2013}, as well as for
multi-objective control \cite{Panagou2013}. %
It was proved in \cite{Tee2009}
that if a BF for a given safe set satisfies Lyapunov-like conditions, then the set
is forward invariant. A BF that is less restrictive in the sense that it is allowed to
decrease when far away from the boundary of the set was proposed in \cite{Aaron2014}. 
Control BFs (CBFs) are 
a version of BFs used for control systems which is employed
to map a state constraint onto a state-feedback control constraint. 
In recent years, CBFs have received increasing attention in the effort to provide \emph{guarantees} for enforcing critical safety constraints
 \cite{Aaron2014} \cite{Glotfelter2017} \cite{xiao2021high} \cite{xiao2023safe1} \cite{li2025robust}.
 {This is largely due to the fact that solving optimal control problems 
 for nonlinear systems with nonlinear constraints
 is computationally prohibitive for real-time applications \cite{li2024safe}. CBFs have been shown to provide a computationally efficient alternative at the expense of potentially more conservative control.
} However, 
this type of CBFs is mainly designed for systems that are initially safe.


To account for systems that initially violate the safety constraint or systems that may evolve outside the safe set due to random events or encounter previously unmodeled unsafe regions (e.g., the traffic control problem for roundabouts studied in \cite{chen2025optimal}),
one may employ robust CBFs by choosing extended class $\mathcal{K}$ functions instead of class $\mathcal{K}$ functions \cite{xu2015robustness} \cite{jankovic2018robust}. In this way, the system state will always be stabilized to the safe set 
whenever the safety constraint is violated since the CBF can be viewed as a general form of Control Lyapunov Function \cite{Aaron2012}. 
However, the use of extended class $\mathcal{K}$ functions cannot ensure that a safety constraint is satisfied in \emph{finite time}. To ensure that the system converges to the safe set in finite time, i.e., ensure finite-time convergence, Control Lyapunov-Barrier functions (CLBFs) have been proposed in \cite{Srinivasan2018} \cite{Li2018}. Nevertheless, this approach does not usually take into account control bounds which may easily conflict with the CLBF constraint, thus leading to infeasibility. Moreover, chattering behaviors at the boundary of the safe set commonly arise as the corresponding derivative
of the extended class $\mathcal{K}$ function
may become infinite at the boundary \cite{xiao2021high2}.  Unifying barrier and Lyapunov functions, as proposed in  \cite{Tee2009},  \cite{Sachan2018}, may also be used in principle, so as to implement finite-time 
convergence
specifications, but this method is only considered for linear state constraints.  
Finite-time convergence to a safe set
that avoids the drawbacks of CLBFs has recently been studied in diffusion models \cite{xiao2023safediffuser} or by a safety recovery approach using time-varying functions and treating safety recovery as a barrier-trajectory design problem \cite{chen2026exacttimesafetyrecoveryusing}, but these works are done without explicitly considering control bounds.

In this paper, we formalize the finite-time convergence method 
proposed in \cite{xiao2023safediffuser} for general control systems that initially violate the safety constraint. We 
propose \emph{finite-time convergent CBFs},
which can strictly ensure the satisfaction of the initially violated constraint within a specified time.
Further, we analyze the feasibility of such finite-time convergent CBFs 
under 
control bounds using reachability analysis and constraint comparison. As a result, we can  
determine parameters for a finite-time convergent CBF such that it will never conflict with control bounds, thus ensuring the feasibility of the eventual optimization problem. Due to the convergence property of the proposed finite-time convergent CBF, we can find the reachable set from the initial system state
and even find the parameters based \emph{only} on  the initial system state that can ensure the feasibility of the finite-time convergent CBF. All these 
properties
facilitate
the parametric design of controllers based on feasibility-guaranteed finite-time convergent CBFs.

In summary, we make the following two contributions in this paper:
\begin{enumerate}
    \item We formalize the finite-time convergent CBF proposed for diffusion models \cite{xiao2023safediffuser} to general 
    nonlinear control affine
     systems, and analyze the associated parametric controller design.
    \item We propose a feasibility analysis method to ensure the existence of a valid finite-time convergent CBF under control bounds using reachability analysis and constraint comparison. 
\end{enumerate}

The remainder of the paper is organized as follows. We present preliminaries on CBFs and CLBFs in Sec.\ref{sec:pre}, and formulate a constrained optimal control problem in Sec.\ref{sec:prob}. We propose the feasibility guaranteed finite-time convergent CBF in Sec. \ref{sec:snc}. A case study of a 2D robot obstacle avoidance problem is given in Sec. \ref{sec:case}.  We conclude this paper in Sec.\ref{sec:conclusion}.

\section{PRELIMINARIES}

\label{sec:pre}

In this section, we introduce preliminary material on CBFs \cite{Aaron2014} for safety and CLBFs for finite-time safety convergence. 
\subsection{Control Barrier Functions}
\begin{definition}
	\label{def:classk} (\textit{Class $\mathcal{K}$ function} \cite{Khalil2002}) A
	continuous function $\alpha:[0,a)\rightarrow[0,\infty), a > 0$
	belongs to class $\mathcal{K}$ if it is strictly increasing and $\alpha(0)=0$. A continuous function $\beta:\mathbb{R}\rightarrow\mathbb{R}$  belongs to extended class $\mathcal{K}$ if it is strictly increasing and $\beta(0)=0$.
\end{definition}

We consider an affine control system (assumed to be deterministic):
\begin{equation}
\dot{\bm{x}}=f(\bm x)+g(\bm x)\bm u \label{eqn:affine}%
\end{equation}
where $\bm x\in X\subset\mathbb{R}^{n}$, $f:\mathbb{R}^{n}\rightarrow\mathbb{R}^{n}$
and $g:\mathbb{R}^{n}\rightarrow\mathbb{R}^{n\times q}$ are locally
Lipschitz continuous, and $\bm u\in U\subset\mathbb{R}^{q}$ is the control constraint set that is defined as follows ($\bm u_{min} \in\mathbb{R}^q,\bm u_{max}\in\mathbb{R}^{q}$):
\begin{equation}
U:=\{\bm u\in\mathbb{R}^{q}:\bm u_{min}\leq\bm u\leq\bm u_{max}\},
\label{eqn:control}%
\end{equation}
where the inequalities in the above are interpreted component-wise.

\begin{definition} (Forward invariance \cite{Aaron2014})
	\label{def:forwardinv} A set $C\subset\mathbb{R}^{n}$ is forward invariant for
	system (\ref{eqn:affine}) if its solutions {for $\bm u\in U$} starting at $\bm x(0) \in C$
	satisfy $\bm x(t)\in C,$ $\forall t\geq0$.
\end{definition}

\begin{definition}
	\label{def:relative} (\textit{Relative degree} \cite{Khalil2002}) The relative degree of a
	(sufficiently many times) differentiable function $b:\mathbb{R}^{n}%
	\rightarrow\mathbb{R}$ with respect to system (\ref{eqn:affine}) is defined as the number
	of times it needs to be differentiated along the dynamics until any component of the control
	$\bm u$ shows up in the corresponding derivative.
\end{definition}

In this paper, we refer to the relative degree of $b$ as the relative
degree of the constraint since the function $b$ is used to define a constraint $b(\bm
x)\geq0$.
For a constraint $b(\bm x)\geq0$ with relative
degree $m \geq 1$, we define $\psi_{0}(\bm
x):=b(\bm x)$, and then we further define a sequence of CBFs $\psi_{i}:\mathbb{R}%
^{n}\rightarrow\mathbb{R},i\in\{1,\dots,m\}$:
\begin{equation}
\begin{aligned} \psi_i(\bm x) := \dot \psi_{i-1}(\bm x) + \alpha_i(\psi_{i-1}(\bm x)),i\in\{1,\dots,m\}, \end{aligned} \label{eqn:functions}%
\end{equation}
where $\alpha_{i},i\in\{1,\dots,m\}$ denote $(m-i)^{th}$ order
differentiable class $\mathcal{K}$ functions.

Next, we define a sequence of safe sets $C_{i}, i\in\{1,\dots,m\}$ corresponding to CBFs (\ref{eqn:functions}):
\begin{equation}
\label{eqn:sets}\begin{aligned} C_i := \{\bm x \in \mathbb{R}^n: \psi_{i-1}(\bm x) \geq 0\}, i\in\{1,\dots,m\}. \end{aligned}
\end{equation}

\begin{definition}
\label{def:hocbf} (\textit{High Order Control Barrier Function (HOCBF)}
\cite{xiao2021high}) Let $C_{i}, i\in\{1\dots,m\}$ be defined by (\ref{eqn:sets}%
) and $\psi_{i}(\bm x), i\in\{1,\dots, m\}$ be defined by
(\ref{eqn:functions}). A function $b: \mathbb{R}^{n}\rightarrow\mathbb{R}$ is
a High Order Control Barrier Function (HOCBF) of relative degree $m$ for
system (\ref{eqn:affine}) if there exist $(m-i)^{th}$ order differentiable
class $\mathcal{K}$ functions $\alpha_{i},i\in\{1,\dots,m-1\}$ and a class
$\mathcal{K}$ function $\alpha_{m}$ s.t.
{\begin{equation}
\label{eqn:constraint}\begin{aligned} \sup_{\bm u\in U}[L_f\psi_{m-1}(\bm x) + L_g\psi_{m-1}(\bm x)\bm u + \alpha_m(\psi_{m-1}(\bm x))] \geq 0, \end{aligned}
\end{equation}
}for all $\bm x\in \cap_{i = 1}^m C_{i}$. In
(\ref{eqn:constraint}), the left part is actually $\psi_m(\bm x)$, $L_{f}$ (or $L_{g}$) denotes Lie derivatives along
$f$ (or $g$).
\end{definition}

The HOCBF in Def. \ref{def:hocbf} is a general form of the relative degree one CBF \cite{Aaron2014},
\cite{Glotfelter2017}. In other words, if $m = 1$, a HOCBF reduces to
the CBF constraint form:
\begin{equation}\label{eqn:cbf0}
L_fb(\bm x) + L_gb(\bm x)\bm u + \alpha_1(b(\bm x))\geq 0,
\end{equation}
 and the HOCBF is also a general form of the exponential CBF
\cite{Nguyen2016} (i.e., setting all class $\mathcal{K}$ functions as linear functions in a HOCBF).

The following theorem establishes the safety guarantee properties of HOCBFs.
\begin{theorem}
\label{thm:hocbf} (\cite{xiao2021high}) Given a HOCBF $b(\bm x)$ from Def.
\ref{def:hocbf} with the safe sets $C_{i}, i\in\{1,\dots, m\}$ defined
by (\ref{eqn:sets}), if $\bm x(0) \in \cap_{i = 1}^m C_{i}$,
then any Lipschitz continuous controller $\bm u(t)\in U$ that satisfies the HOCBF constraint in
(\ref{eqn:constraint}), $\forall t\geq0$ renders $\cap_{i = 1}^m C_{i}$ forward invariant for system (\ref{eqn:affine}).
\end{theorem}

\subsection{Control Lyapunov-Barrier Function}
\begin{definition}
\label{def:clbf} (\textit{Control Lyapunov-Barrier Function (CLBF)})  A function $b: \mathbb{R}^{n}\rightarrow\mathbb{R}$ is
a Control Lyapunov-Barrier Function (CLBF) for
system (\ref{eqn:affine}) if there exist a
$p > 0$ and $0<q<1$ such that
{\begin{equation}
\label{eqn:clbf}\begin{aligned} \sup_{\bm u\in U}[L_fb(\bm x) + L_gb(\bm x)\bm u + pb^q(\bm x)] \geq 0, \end{aligned}
\end{equation}
}for all $\bm x\notin C_1$.
\end{definition}

CLBFs are mainly designed for systems that initially violate a safety constraint, and they ensure that the constraint is satisfied within a specified time.
In particular, a CLBF guarantees that the safety constraint $b(\bm x)\geq 0$ is satisfied within time $\frac{b^{1-q}(\bm x(0))}{p(1-q)}$. A class of such high-order CLBFs has been studied in \cite{xiao2021high2}.

In the literature, constrained optimal control problems with quadratic cost functions are commonly reformulated \cite{Aaron2014}
by incorporating CBFs or HOCBFs to replace original state constraints.
The time line is discretized and an optimization problem with constraints given by the CBFs/HOCBFs is solved at each time step. 
Note that these constraints are linear in control 
(see (\ref{eqn:constraint}))
and the state value is given and fixed at the beginning of each time interval, therefore, the resulting optimization problem becomes a Quadratic
Program (QP). The optimal control obtained from each QP is then applied over the entire time interval. The system state is
updated by integrating dynamics (\ref{eqn:affine}). This procedure is repeated until the final time is reached. Mathematically, suppose the objective is to minimize $\int_0^Tu^T H \bm u$ 
for system (\ref{eqn:affine}), where $H\in\mathbb{R}^{q\times q}$ is positive definite. We 
	partition the time interval $[0,T]$  into a set of small time intervals $\{[0, \Delta t), [\Delta t,2\Delta t),\dots\}$, where $\Delta t > 0$. Within each $[\omega \Delta t, (\omega+1) \Delta t)$ ($\omega = 0,1,2,\dots$), the control input is assumed to be constant.
	Then {at $t = \omega \Delta t$,} the following QP is solved:
	\begin{equation} \label{eqn:obj}
	\begin{aligned}
	\min_{\bm u(t)} & \bm u^T(t) H \bm u(t) \\
	&\text{s.t. }\bm u_{min}\leq\bm u\leq\bm u_{max},\\
	& L_f\psi_{m-1}(\bm x) + L_g\psi_{m-1}(\bm x)\bm u + \alpha_m(\psi_{m-1}(\bm x)) \geq 0.
	\end{aligned}
	\end{equation}

CLBFs can also be incorporated into the above QP to ensure that the system state satisfies the safety constraint within a specified time. However, existing works usually do not consider the control bound as it can easily conflict with the CLBF constraint. Moreover, the CLBF may exhibit chattering behavior at the boundary of the safe set, as the derivative of $pb^q(\bm x)$ 
in (\ref{eqn:clbf})
goes to infinity when $b(\bm x) = 0$.  In what follows, we show how we can ensure the feasibility of the finite-time convergence requirement while avoiding the chattering issue with a new proposed finite-time convergent CBF.

\section{PROBLEM FORMULATION AND APPROACH}
\label{sec:prob}

We begin by formulating a general constrained optimal control problem as follows.

$\textbf{Objective}$: (Minimizing cost) We consider an optimal control problem for system (\ref{eqn:affine}) with the cost defined as:
\begin{equation}\label{eqn:cost}
\min_{\bm u(t)}\int_{0}^{T}\mathcal{C}(||\bm u(t)||)dt,
\end{equation}
where $T > 0$, $||\cdot||$ denotes the 2-norm of a vector, $\mathcal{C}(\cdot)$ is a strictly increasing function of its argument.




$\textbf{Safety guarantees}$: 
System (\ref{eqn:affine}) should always satisfy a safety constraint:
\begin{equation} \label{eqn:safetycons}
b(\bm x(t))\geq 0,~~ \forall t\in[0,T],
\end{equation}
where $b: \mathbb{R}^n\rightarrow\mathbb{R}$ is continuously differentiable and has relative degree $m\in\mathbb{N}$ with respect to system (\ref{eqn:affine}). The problem can be extended to a more general setting with multiple simultaneous safety constraints of this form.


$\textbf{Safe Goal Constraint}$: 
System (\ref{eqn:affine}) should satisfy an initially violated safety constraint at time $t_f \in [0, T]$:
\begin{equation} \label{eqn:safetyconv}
h(\bm x(t_f))\geq 0,
\end{equation}
where $h: \mathbb{R}^n\rightarrow\mathbb{R}$ is continuously differentiable and $h(\bm x(0)) < 0$.

$\textbf{Control constraints}$: The control of the system (\ref{eqn:affine}) 
should always satisfy the control bound as defined in (\ref{eqn:control}).

A control policy is \textit{feasible} for system (\ref{eqn:affine}) if all the safety constraints (\ref{eqn:safetyconv}) and (\ref{eqn:safetycons}) and control bound (\ref{eqn:control}) are satisfied for all times.

\begin{problem}\label{prob:general}
	Find a \textit{feasible} control policy for system (\ref{eqn:affine}) such that the cost (\ref{eqn:cost}) is minimized, and all safety constraints \eqref{eqn:safetyconv} and \eqref{eqn:safetycons} are satisfied.
\end{problem}

\textbf{Approach:} To solve Problem \ref{prob:general}, our method is based on the proposed finite-time convergent CBF that enforces the constraint (\ref{eqn:safetyconv}). We use a CBF/HOCBF to enforce the safety constraint (\ref{eqn:safetycons}), and directly take the  cost (\ref{eqn:cost}) as the objective to minimize. Our proposed finite-time convergent CBF is flexible to be designed in a way that it will never conflict with the control bounds using constraint comparison and reachability analysis, as described next, thus ensuring the feasibility of the safety convergence problem.

\section{Feasibility-Guaranteed Finite-Time convergent CBFs}
\label{sec:snc}

In this section, we define a new type of CBF termed \emph{finite-time convergent CBF} with feasibility guarantees. This finite-time convergent CBF was first studied in diffusion models \cite{xiao2023safediffuser} without considering any control bounds.

\subsection{Finite-Time convergent CBFs}

Consider a safety constraint $h(\bm x)\geq 0$ that has relative degree one 
with respect to (\ref{eqn:affine}) and is initially violated as shown in (\ref{eqn:safetyconv}) 
(the case where the relative degree is $>1$ is the subject of ongoing research).
The constraint has to be satisfied at time $t_f \in [0, T]$. 
Te first problem we address is that employing a CLBF to enforce the finite-time convergence requirement may introduce a chattering behavior at the boundary of the safe set.
To avoid this problem, we propose a new finite-time convergent CBF.

It is known that a CBF with extended class $\mathcal{K}$ functions possesses the Lyapunov stability property if the corresponding safety constraint is initially violated \cite{xiao2021high2}. 
Although this property is helpful to make the system state converge to the safe set, the system state may never reach the safe set 
by a desired finite time
as this convergence is asymptotic.

To employ the asymptotic stability of a CBF, we begin by strengthening the constraint with a constant $r>0$ by defining a new constraint 
\begin{equation} \label{eqn:str}
    s(\bm x) := h(\bm x) - r,
\end{equation}

Then, we define $s(\bm x)$ as a \emph{valid} finite-time convergent CBF as follows:
\begin{definition}  [Valid Finite-Time Convergent CBF]\label{def:ftcbf}
    Given a safety convergence requirement $h(\bm x) \geq 0$  with $h(\bm x(0)) < 0$ as in (\ref{eqn:safetyconv}),
    a strengthened safety convergence function $s(\bm x)$ defined in (\ref{eqn:str}) is a valid finite-time convergent CBF if 
    \begin{equation} \label{eqn:ftcbf}
        \sup_{\bm u\in U}\left[L_fs(\bm x) + L_gs(\bm x)\bm u + ks(\bm x)\right] \geq 0,
    \end{equation}
    for all $\bm x\in X$, where $k > 0$.
\end{definition}

The following theorem shows the effectiveness of a 
valid
finite-time convergent CBF:
\begin{theorem} \label{thm:ftcbf}
    Given a valid finite-time convergent CBF $s(\bm x)$ defined as in (\ref{def:ftcbf}), if the parameters $r$ and $k$ are chosen such that 
    \begin{equation}
        r \geq (r - h(\bm x(0))) e^{-kt_f},
    \end{equation}
    then the system state converges to the safe set defined by $h(\bm x)\geq 0$ in finite time $t_f$, i.e., $h(\bm x(t_f))\geq 0$.
\end{theorem}
\begin{proof}
Define $V(\bm x)$ as
\begin{equation} \label{eqn:0}
    V(\bm x) := -s(\bm x) =  r - h(\bm x).
\end{equation}
Since $r > 0$ and $h(\bm x(0)) < 0$, we have
$V(\bm x(0)) > 0$.

Substituting equation \eqref{eqn:0} into the constraint in (\ref{eqn:ftcbf}), we have 
\begin{equation}
    L_fV(\bm x) + L_gV(\bm x)\bm u + k V(\bm x) \leq 0,
\end{equation}
which is equivalent to
\begin{equation} \label{eqn:1}
    \dot V(\bm x) + kV(\bm x) \leq 0.
\end{equation}
\noindent Now, let us consider the case where the equality holds, i.e.,
\begin{equation} 
    \dot V(\bm x) + kV(\bm x) = 0,
\end{equation}
the solution of $V(\bm x)$ is then given by
\begin{equation} \label{eqn:2}
    V(\bm x(t)) = V(\bm x(0))e^{-kt}.
\end{equation}
\noindent Applying the comparison lemma \cite{Khalil2002}, equations (\ref{eqn:1})-(\ref{eqn:2})  imply
\begin{equation} \label{eqn:V_ineq}
    V(\bm x(t)) \leq V(\bm x(0)) e^{-kt}
\end{equation}
Substituting (\ref{eqn:0}) into \eqref{eqn:V_ineq}, we have 
\begin{equation*}
    r - h(\bm x(t)) \leq (r - h(\bm x(0))) e^{-kt},
\end{equation*}
which can be rewritten as
\begin{equation}\label{eqn:r_cond}
     - h(\bm x(t)) \leq (r - h(\bm x(0))) e^{-kt} - r
\end{equation}
Since the parameters $r$ and $k$ are chosen such that $r \geq (r - h(\bm x(0))) e^{-kt_f}$, \eqref{eqn:r_cond} yields
\begin{equation}
     - h(\bm x(t_f)) \leq (r - h(\bm x(0))) e^{-kt_f} - r \leq 0,
\end{equation}
Thus, we have $h(\bm x(t_f)) \geq 0$, and the system state converges to the safe set defined by $h(\bm x)\geq 0$ in finite time $t_f$.
\end{proof}

Note that \eqref{eqn:V_ineq} implies that $V(\bm x)=r-h(\bm x)$ asymptotically converges to 0, indicating that $h(\bm x)$ asymptotically converges to $r>0$, thereby avoiding the chattering issues that may happen in the CLBF method.


\begin{remark} [Existence of a finite-time convergent CBF]
As shown in Def. \ref{def:ftcbf}, finding a valid finite-time convergent CBF remains challenging due to the control bound in (\ref{eqn:control}). Specifically, a valid finite-time convergent CBF should satisfy both the corresponding constraint in (\ref{eqn:ftcbf}) and the control bound (\ref{eqn:control}) simultaneously. This 
problem remains largely underexplored in the literature and it is the second problem we address in the reminder of this paper by studying the feasibility of the proposed finite-time convergent CBFs.
\end{remark}


\subsection{Feasibility Analysis of Finite-Time convergent CBFs}

In this subsection, we show how to find a valid finite-time convergent CBF using constraint comparison and reachability analysis.

\textbf{Reachable Set.} As shown in the proof of Thm. \ref{thm:ftcbf}, the inequality \eqref{eqn:V_ineq} holds for all $\bm x$, indicating $V(\bm x) = -s(\bm x) = r - h(\bm x)$ is strictly decreasing. Since $r$ is a positive constant, $h(\bm x)$ is strictly increasing. 
Therefore, we can find the reachable set $R(\bm x(0))$ for system state $\bm x$ in a finite-time convergent CBF:
\begin{equation}\label{eqn:reachability}
    R(\bm x(0)) = \{\bm y\in\mathbb{R}^n: h(\bm x(0))\leq h(\bm y) \leq 0\}.
\end{equation}
From equation \eqref{eqn:reachability}, the reachable set $R(\bm x(0))$ is closed if the safe set defined by $h(\bm x)\geq 0$ is closed.

\textbf{Constraint Comparison.} Suppose $L_gs(\bm x) \leq \bm 0$ (the inequality is interpreted component-wise) and all the components in $L_gs(\bm x)$ do not change sign for all $\bm x\in R(\bm x(0))$. The cases in which those assumptions are not satisfied can be addressed similarly, as shown in \cite{Xiao2021}. 

Since $L_gs(\bm x) \leq \bm 0$, we multiply $-L_gs(\bm x)$ with the control bound (\ref{eqn:control}) and get
\begin{equation} \label{eqn:enlarge}
    -L_gs(\bm x)\bm u_{min} \leq -L_gs(\bm x)\bm u\leq -L_gs(\bm x)\bm u_{max},
\end{equation}

The finite-time convergent CBF constraint in (\ref{eqn:ftcbf}) can be rewritten as
\begin{equation} \label{eqn:ftcbf0}
    -L_gs(\bm x)\bm u \leq L_fs(\bm x) + ks(\bm x)
\end{equation}

Comparing the control constraint (\ref{eqn:enlarge}) with the finite-time convergent CBF constraint (\ref{eqn:ftcbf0}), their intersection is non-empty if 
\begin{equation*}
    L_fs(\bm x) + ks(\bm x) \geq -L_gs(\bm x)\bm u_{min}.
\end{equation*}

We define a \emph{candidate} finite-time convergent CBF as follows: 
\begin{definition} [candidate finite-time convergent CBF]\label{def:ftcbf-can}
    Given a safety convergence requirement $h(\bm x) \geq 0$  with $h(\bm x(0)) < 0$ as in (\ref{eqn:safetyconv}),
    a strengthened safety convergence function $s(\bm x)$ defined in (\ref{eqn:str}) is a candidate finite-time convergent CBF if there exists a controller $\bm u$ such that
    \begin{equation} \label{eqn:ftcbf-can}
        L_fs(\bm x) + L_gs(\bm x)\bm u + ks(\bm x) \geq 0,
    \end{equation}
    for all $\bm x\in X$, where $k > 0$.
\end{definition}
The main difference between a valid finite-time convergent CBF in Def. \ref{def:ftcbf} and a candidate finite-time convergent CBF in Def. \ref{def:ftcbf-can} lies in that a valid finite-time convergent CBF considers the control bound $u\in U$ given in (\ref{eqn:control}) while a candidate valid finite-time convergent CBF does not.

The following lemma will be used to show the feasibility of the finite-time convergent CBF:
\begin{lemma} \label{lem:fea}\cite{Xiao2021}
If the control $\bm u$ is such that the finite-time convergence constraint in (\ref{eqn:ftcbf0}) is conflict free with (\ref{eqn:enlarge}), then the control bound (\ref{eqn:control}) is also conflict free with the the finite-time convergence constraint in (\ref{eqn:ftcbf0}).
\end{lemma}

The following theorem shows the existence of a valid finite-time convergent CBF (i.e., feasibility guaranteed finite-time convergent CBF):
\begin{theorem} \label{thm:fea}
    Given a candidate finite-time convergent CBF $s(\bm x)$ as in Def. \ref{def:ftcbf-can} with $r$ and $k$ chosen such that $r \geq (r - h(\bm x(0))) e^{-kt_f}$, if the parameters $r$ and $k$ satisfy \begin{equation}\label{eqn:thm_con}
        \min_{\bm x\in R(\bm x(0))} L_fs(\bm x) + ks(\bm x) \geq \max_{\bm x\in R(\bm x(0))} -L_gs(\bm x)\bm u_{min},
    \end{equation} 
    then, $s(\bm x)$ is a valid finite-time convergent CBF defined in Def. \ref{def:ftcbf}, and the system state is guaranteed to satisfy $h(\bm x(t_f))\geq 0$ under the control bound (\ref{eqn:control}).
\end{theorem}
\begin{proof}
Since $$\min_{\bm x\in R(\bm x(0))} L_fs(\bm x) + ks(\bm x) \geq \max_{\bm x\in R(\bm x(0))} -L_gs(\bm x)\bm u_{min},$$ we have 
\begin{equation}\label{eqn:can_ftcbf}
    L_fs(\bm x) + ks(\bm x) \geq -L_gs(\bm x)\bm u_{min}
\end{equation}
satisfied for all $\bm x\in R(\bm x(0))$ as $R(\bm x(0))$ is the reachable set of system states under the candidate finite-time convergent CBF.
Since $s(\bm x)$ is a candidate finite-time convergent CBF, it follows from \eqref{eqn:ftcbf-can} that the following inequality holds for all $\bm x$:
\begin{equation}\label{ineq:ftcbf-can}
    L_fs(\bm x) + ks(\bm x) \geq -L_gs(\bm x)\bm u
\end{equation}

Comparing \eqref{eqn:enlarge} and \eqref{ineq:ftcbf-can}, the satisfaction of \eqref{eqn:can_ftcbf} implies that the constraint
$$-L_gs(\bm x)\bm u_{min} \leq -L_gs(\bm x)\bm u\leq -L_gs(\bm x)\bm u_{max},
$$
is conflict-free with the candidate finite-time convergent CBF constraint in Def. \ref{def:ftcbf-can}:
$$ L_fs(\bm x) + L_gs(\bm x)\bm u + ks(\bm x) \geq 0.
$$

By Lemma \ref{lem:fea}, the finite-time convergent CBF constraint in Def. \ref{def:ftcbf-can} is also conflict-free with the control bound (\ref{eqn:control}), and we have that 
$$
 \sup_{\bm u\in U}\left[L_fs(\bm x) + L_gs(\bm x)\bm u + ks(\bm x)\right] \geq 0,  
    ~~\forall \bm x,
$$
Hence, $s(\bm x)$ is a valid finite-time convergent CBF. Since $r$ and $k$ are chosen such that $r \geq (r - h(\bm x(0))) e^{-kt_f}$, by Thm. \ref{thm:ftcbf} we have that $h(\bm x(t_f))\geq 0$ is guaranteed to be satisfied under the control bound (\ref{eqn:control}).
\end{proof}

Note that if $L_g s(\bm x)>0$, then the condition \eqref{eqn:thm_con} in Thm. \ref{thm:fea} becomes 
\begin{equation*}
        \min_{\bm x\in R(\bm x(0))} L_fs(\bm x) + ks(\bm x) \geq \max_{\bm x\in R(\bm x(0))} -L_gs(\bm x)\bm u_{max}.
    \end{equation*} 

\begin{remark} [Choosing $r$ and $k$]
In practice, since $s(\bm x)$ is strictly increasing, as shown in the proof of Thm. \ref{thm:ftcbf}, it suffices to check the condition $L_fs(\bm x(0)) + ks(\bm x(0)) \geq -L_gs(\bm x(0))\bm u_{min}$ at the initial time 0, especially when $L_gs(\bm x)$ is a constant and when $L_fs(\bm x)$ is either a constant or increasing. 
This happens in many examples, such as adaptive cruise control and 2D robot navigation control examples (considered in the next section of this paper).  This observation significantly facilitates the choice of parameters $r$ and $k$ in order to find a valid finite-time convergent CBF without considering the entire reachable set of the system.
\end{remark}

\begin{remark} (Addressing conflict between safety constraint (\ref{eqn:safetycons}) and safety convergence (\ref{eqn:safetyconv}))
    We can employ a similar technique of finding a valid finite-time convergent CBF to address the possible conflict between safety constraint (\ref{eqn:safetycons}) and safety convergence (\ref{eqn:safetyconv}). In practice, however, we find that the main feasibility issue for a finite-time convergent CBF lies in the conflict between the safety convergence (\ref{eqn:safetyconv}) and the control bound (\ref{eqn:control}). 
\end{remark}

\textbf{Solution to Problem \ref{prob:general}.} To address Problem \ref{prob:general}, we use a finite-time convergent CBF to enforce the safety convergence (\ref{eqn:safetyconv}), and use a CBF/HOCBF to enforce the safety constraint(s) (\ref{eqn:safetycons}). Then, we can formulate an optimization problem, and use the time discretization method introduced at the end of Sec. \ref{sec:pre} to solve it. Eventually, we get a sequence of QPs to be solved at discretized time instants.

\section{CASE STUDY
}

\label{sec:case}

In this section, we consider a robot 2D obstacle avoidance example, and compare our proposed method with a CLBF-based controller. We use the \textit{Quadprog} to solve the QP, and use the \textit{ODE45} to integrate the dynamics in \textit{MATLAB}. All the computation runs on a {\it Intel(R) Core(TM) i7-14700HX CPU} computer. The computation time for each QP is less than $0.01s$.

We consider single integrator dynamics $\dot x = u_1, \dot y = u_2$ for the robot,
where $(x,y)\in\mathbb{R}^2$ is the location of the robot, and $\bm u = (u_1,u_2)\in\mathbb{R}^2$ is the control. The \textbf{control bounds} are defined as
\begin{equation} \label{case:bound}
 \begin{aligned}
u_{1,min}\leq u_1\leq u_{1,max},\\
u_{2,min}\leq u_2\leq u_{2,max},
\end{aligned}
\end{equation}
where $u_{1,min} < 0, u_{2,min} < 0, u_{1,max} > 0, u_{2,max} > 0$. $\bm u_{max} = (u_{1,max}, u_{2,max}), \bm u_{min} = (u_{1,min}, u_{2,min})$.

We consider four obstacles in the environment, and the corresponding \textbf{safety constraints} are defined as:
\begin{equation}\label{eqn:safetycons-robot}
    b_i(x):=(x - x_{o,i})^2 + (y - y_{o,i})^2 - R_i^2, i\in\{1,\dots,4\}
\end{equation}
where $(x_{o,i}, y_{o,i})\in\mathbb{R}^2$ denotes the location of the obstacle $i$, and $R_i > 0$ denotes its radius.

The \textbf{safe goal constraint} is defined as
\begin{equation} \label{eqn:exam-conv}
    (x - x_{c})^2 + (y - y_{c})^2 \leq R^2,
\end{equation}
where $(x_{c}, y_{c})\in\mathbb{R}^2$ and $R > 0$, and $h(\bm x) =  R^2 - (x - x_{c})^2 - (y - y_{c})^2\geq 0$ with $h(\bm x(0)) < 0$. We require $h(\bm x(t_f)) \geq 0$ for a finite time $t_f > 0$.

 The \textbf{objective function} (\ref{eqn:cost}) is explicitly defined as 
 \begin{equation} \label{case:obj}
     \min_{\bm u(t)}\int_{0}^{T}||\bm u(t)||^2dt.
 \end{equation}

The problem is to find an optimal control that minimizes (\ref{case:obj}) subject to (\ref{case:bound}) - (\ref{eqn:exam-conv}).
 
 We use both a CLBF \cite{chen2026exacttimesafetyrecoveryusing} and a finite-time convergent CBF to enforce the safety convergence requirement (\ref{eqn:exam-conv}), i.e., $h(\bm x)\geq 0$. In the case of using a CLBF, we choose $q = 1/3$ in Def. \ref{def:clbf} and $p = \frac{h^{1-q}(\bm x(0))}{t_f(1-q)}$.
 
 For the case of a finite-time convergent CBF, we define a strengthened constraint
 \begin{equation}
     s(\bm x) = h(\bm x) - r,
 \end{equation}
 We then define $s(\bm x)$ as a \emph{candidate} finite-time convergent CBF as in Def. \ref{def:ftcbf-can}. In order to ensure $h(\bm x(t_f))\geq 0$, we choose $r$ and $k$ such that 
\begin{equation}\label{eqn:ft-exam}
 r \geq (r - h(\bm x(0))) e^{-kt_f}
 \end{equation}
 following Thm. \ref{thm:ftcbf}. Further, to ensure a \emph{valid} finite-time convergent CBF, we require 
 \begin{equation}\label{eqn:fea-exam}
 L_fs(\bm x) + ks(\bm x) \geq -L_gs(\bm x)\bm u_{max}, \forall \bm x
 \end{equation}
 following Thm. \ref{thm:fea}. Note that in this case $L_gs(\bm x) > 0$, thus we consider $\bm u_{max}$ 
 in (\ref{eqn:thm_con}) instead of $\bm u_{min}$. As $L_fs(\bm x) = 0$ and $-L_gs(\bm x)\bm u_{max}$ has the maximum value at the initial state, we only need to consider the condition in Thm. \ref{thm:fea}  at the initial time 0, and (\ref{eqn:fea-exam}) becomes:
 $$
 k \leq \frac{-L_gs(\bm x(0))\bm u_{max}}{h(\bm x(0)) - r}.
 $$
 Combining the last equation with the finite-time convergence condition (\ref{eqn:ft-exam}), it follows that $r, k$ have to satisfy:
 $$
 -\frac{1}{t_f}ln\left(\frac{r}{r - h(\bm x(0))}\right)\leq k \leq \frac{-L_gs(\bm x(0))\bm u_{max}}{h(\bm x(0)) - r}.
 $$
 Finally, we use CBFs to enforce all the safety constraints, and we get the following QP to be solved 
 at each time step following time discretization:
\begin{equation}
\begin{aligned}
    &\min_{\bm u(t)}||\bm u(t)||^2 \\
    &\text{s.t.} \\
    &L_fs(\bm x) + L_gs(\bm x)\bm u + ks(\bm x) \geq 0, \\
    &L_fb_i(\bm x) + L_gb_i(\bm x)\bm u + \alpha_i(b_i(\bm x)) \geq 0, i\in\{1,\dots, 4\},\\
    &\bm u_{min}\leq\bm u\leq\bm u_{max}.
    \end{aligned}
\end{equation} 
where $b_i(\bm x) = (x - x_{o,i})^2 + (y - y_{o,i})^2 - R_i^2$.

\begin{figure}[t]
	\centering
	\includegraphics[scale=0.6]{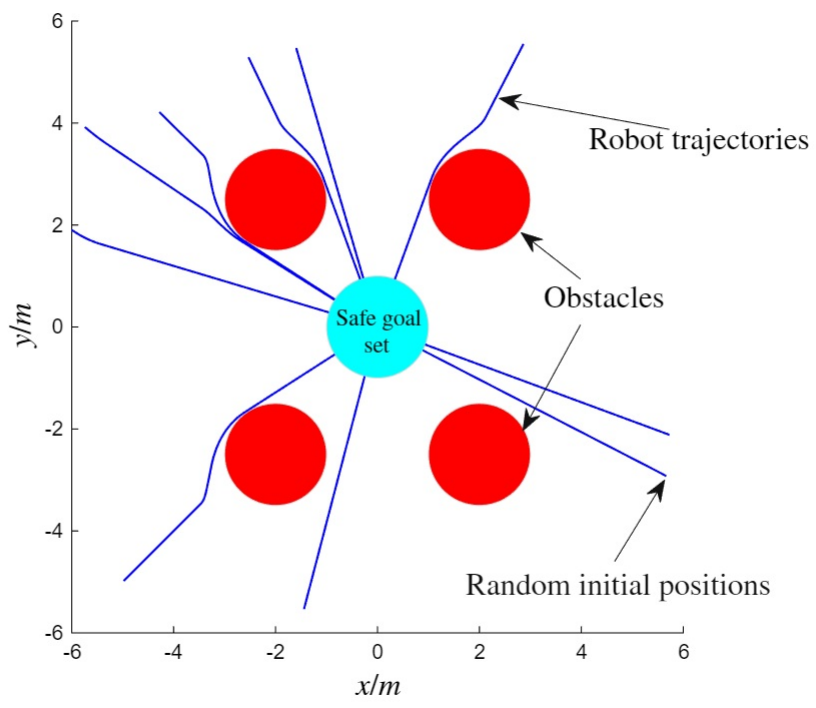}
	\caption{Robot trajectories under different random initial positions with the proposed finite-time convergent CBF.  The robot can reach the safety convergence set in specified time $t_f = 6s$ while guaranteeing safety (avoidance with respect to all obstacles).	}
	\label{fig:traj}
\end{figure}

\begin{figure}[htbp]
	\centering
	\includegraphics[scale=0.49]{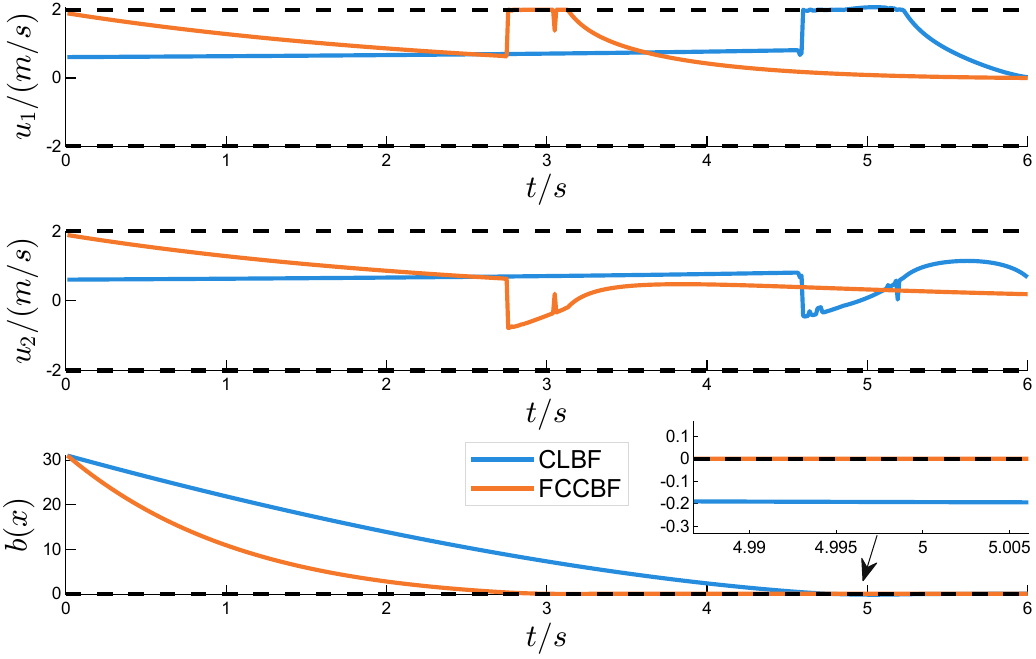}
	\caption{Control and $b(\bm x)$ profiles under CLBF and the Finite-time convergent CBF (FCCBF) method. The control bounds are strictly satisfied. The safety constraint $b(x) \geq 0$ is not satisfied at around $t=5s$ under the CLBF method, whereas the proposed method remains nonnegative for all time.
	}
	\label{fig:control-clbf}
\end{figure}

\textbf{Simulation parameters.} We define all the class $\mathcal{K}$ functions as linear functions with slope 2 in CBFs. The initial position $\bm x(0)$ is randomly generated.  In Fig. \ref{fig:traj}, the center of safe goal set 
is $(x_c, y_c) = (0, 0)m,$ the center of the four obstacles are $ (x_{o,1}, y_{o,1}) = (2, 2.5)m,$ $(x_{o,2}, y_{o,2}) = (-2, 2.5)m,$ $(x_{o,3}, y_{o,3}) = (2, -2.5)m,$ $(x_{o,4}, y_{o,4}) = (-2, -2.5)m$. In Fig. \ref{fig:control-clbf}, the center of the obstacle is $(x_{o}, y_{o}) = (-1, -1)m$. Other parameters are $R_i = 1m, R = 1m, u_{1,max} = -u_{1,min} = 2 m/s,$ $u_{2,max} = -u_{2, min} = 2m/s$,  $t_f = 6s$.

\begin{figure}[htbp]
	\centering
	\includegraphics[scale=0.49]{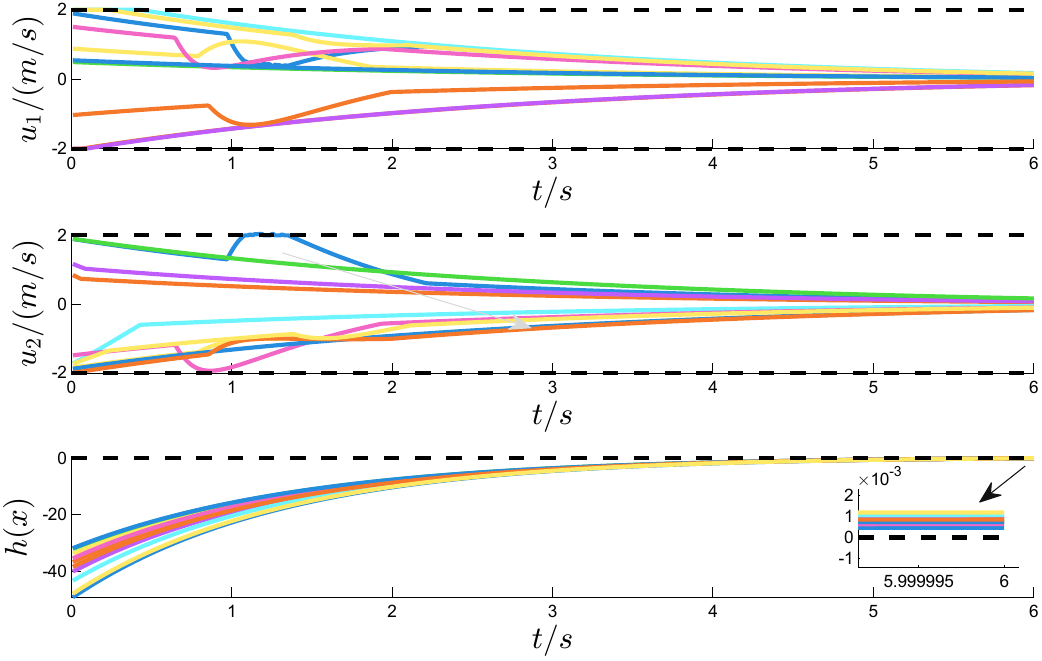}
	\caption{Control and $h(\bm x)$ profiles corresponding to the 10 trajectories in Fig. \ref{fig:traj} with the proposed feasibility guaranteed finite-time CBF. The control bounds are strictly satisfied, and the finite-time convergence $h(\bm x(t_f)) \geq 0$ is satisfied at $t_f = 6s$.
	}
	\label{fig:control}
\end{figure}

\begin{figure}[htbp]
	\centering
	\includegraphics[scale=0.49]{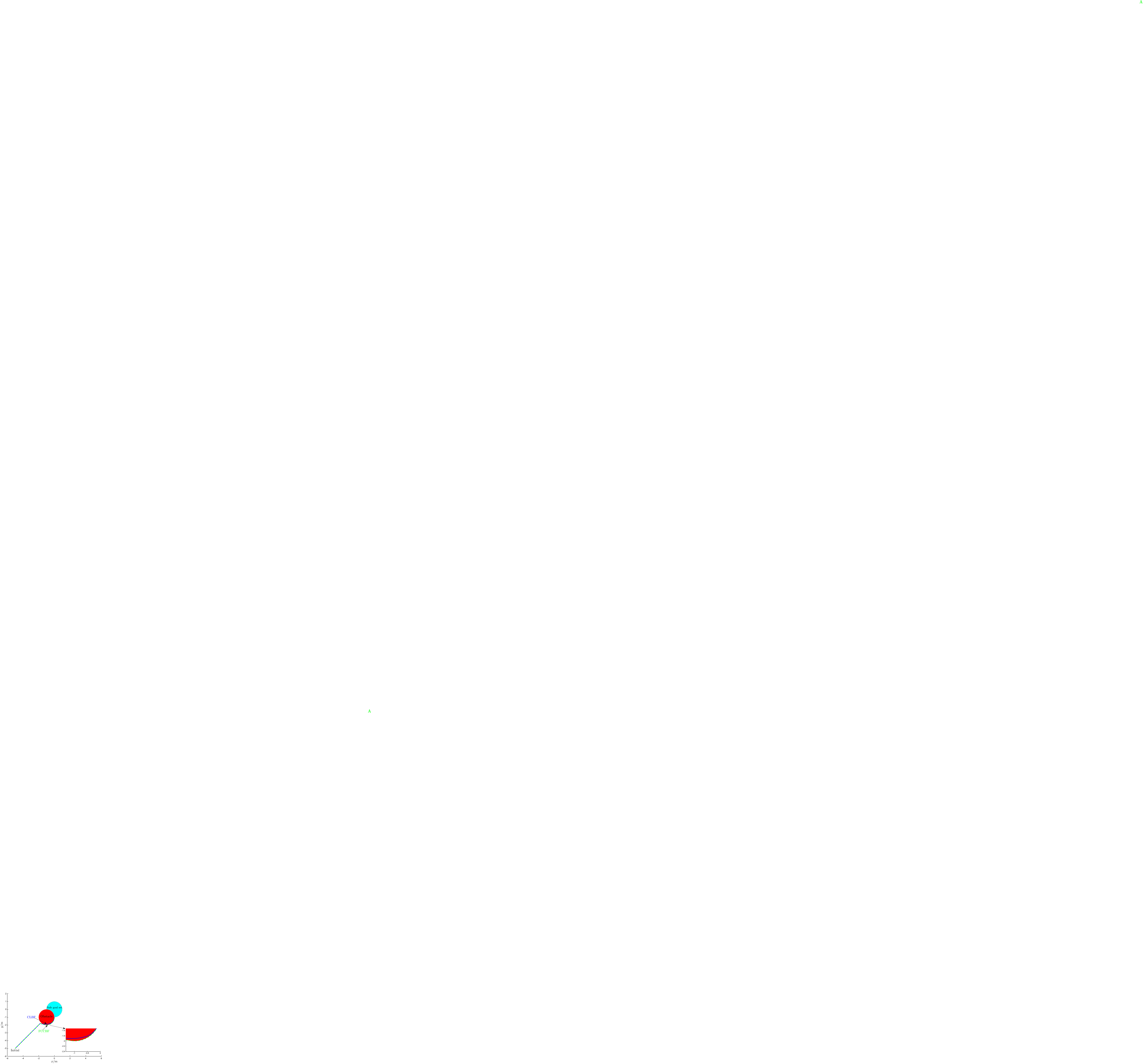}
	\caption{Robot trajectories under the proposed Finite-time convergent CBF (FCCBF) and CLBF method. The safety constraint \eqref{eqn:safetycons-robot} is violated as the blue trajectory (CLBF) enters the red circle, indicating a collision with the obstacle. The green curve (FCCBF) remains outside of the obstacle, ensuring conflict-free all the time. 
	}
	\label{fig:traj-clbf}
\end{figure}

\textbf{Results.} The simulation results are summarized in Figs. \ref{fig:traj} - \ref{fig:control-clbf}. As shown in Fig. \ref{fig:traj}, where the  
safe goal constraint 
$h(x(t_f))\geq 0$ is illustrated by the blue circle, indicating that the robot
is required to converge
to this set in finite time.
The red circles represent obstacles corresponding to unsafe sets, which the robot must avoid at all times. The initial positions of robots are randomly generated. Using the proposed feasibility guaranteed finite-time convergent CBF method, we can ensure that the robot 
reach the desired safe set (blue circle) while avoiding all the obstacles. 
Specifically, set $t_f=6s$, as shown in the last frame of Fig. \ref{fig:control}, the robot under 10 randomly generated initial positions ultimately
reaches the safe set within finite time $t_f$ as $h_i(x(t_f=6))>0$ for $i=1,...,10$, and the control bounds are also strictly satisfied for all the trajectories as shown in the top two plots in Fig. \ref{fig:control}. In the bottom plot, the finite-time convergent CBFs $h_i(x), i=1,...,10$ increase asymptotically to the desired value $r$, thereby demonstrateing the asymptotic convergence property proved in Thm. \ref{thm:ftcbf}. 


Moreover, we compare the proposed finite-time convergent CBF with the CLBF method. The resulting robot trajectories under the two methods are shown in Fig. \ref{fig:traj-clbf}, 
where the blue curve corresponds to the CLBF method and the green curve represents the proposed method
 labeled FCCBF.
It can be observed that the blue trajectory enters the red circle, indicating a collision with the obstacle, whereas the green trajectory remains outside the red circle, demonstrating that the proposed finite-time convergent CBF ensures the feasibility of the control policy while maintaining safety. 

Specifically, the control inputs of the proposed finite-time convergent CBF and CLBF methods, along with the evolution of the safety constraint $b(x)$, are shown in Fig. \ref{fig:control-clbf}, where the blue curve corresponds to the CLBF method and the orange curve represents the proposed finite-time convergent CBF (FCCBF) approach. 
The top two plots illustrate the control policies, where the control bounds are strictly satisfied for both methods. The sudden change in each curve indicates that the robot is avoiding 
the obstacle, which means the robot under CLBF gets close to the obstacle slower than the proposed FCCBF method. 
As a result, in the bottom plot, the blue curve becomes negative around $t = 5 \mathrm{s}$, indicating a violation of the safety constraint, which means the corresponding CLBF control policy is infeasible. 
The reason behind this is that 
the time left for the CLBF-based controller to react to the obstacle is less than the time available using the proposed method,
which makes the safety constraint conflict with the control bound.
Meanwhile, the orange curve is always above 0, demonstrating that the proposed method guarantees safety and also maintains feasibility.

\section{CONCLUSION \& FUTURE WORK}

\label{sec:conclusion}

This paper proposes a feasibility guaranteed finite-time convergent CBF that can ensure the system to satisfy an initially violated constraint under control bound, and this is achieved by employing the asymptotic property of the strengthened constraint and reachability analysis with constraint comparison. We validate the effectiveness of the proposed methods on a 2D obstacle avoidance example with results showing finite-time convergence with feasibility guarantees. Future work will focus on the extension of the framework to high-order systems, as well as how to address the possible conflict between safety constraint and safety convergence requirements.






\bibliographystyle{IEEEtran}
\bibliography{CBF}

\end{document}